\documentclass[showpacs,twocolumn,aps,preprintnumbers,letterpaper]{revtex4}
\usepackage{amsmath,amssymb}
\usepackage{epsfig}
\usepackage{graphicx}
\usepackage{amsmath}
\usepackage{slashed}
\usepackage{amsfonts}
\usepackage{epstopdf}
\usepackage{color}

\newcommand{\be}{\begin{equation}}
\newcommand{\ee}{\end{equation}}
\newcommand{\bear}{\begin{eqnarray}}
\newcommand{\eear}{\end{eqnarray}}
\newcommand{\ba}{\begin{array}}
\newcommand{\ea}{\end{array}}

\def\be{\begin{eqnarray}}
\def\ee{\end{eqnarray}}
\def\bea{\be}
\def\eea{\ee}

\def\roughly#1{\mathrel{\raise.3ex\hbox{$#1$\kern-.75em%
\lower1ex\hbox{$\sim$}}}}

\definecolor{davecolor}{rgb}{0.95,  0.5,  0.2}

\def\({\left(}
\def\){\right)}
\def\[{\left[}
\def\]{\right]}
\def\<{\langle}
\def\>{\rangle}

\newcommand{\bwt}{\begin{widetext}}
\newcommand{\ewt}{\end{widetext}}

\newcommand{\bi}{\begin{itemize}}
\newcommand{\ei}{\end{itemize}}
\newcommand{\ben}{\begin{enumerate}}
\newcommand{\een}{\end{enumerate}}
\newcommand{\bca}{\begin{cases}}
\newcommand{\eca}{\end{cases}}
\newcommand{\bln}{\begin{align}}
\newcommand{\eln}{\end{align}}
\newcommand{\bst}{\begin{split}}
\newcommand{\est}{\end{split}}

  \long\def\comment#1{ }

  \newcommand{\beq}{\begin{eqnarray}}
  \newcommand{\eeq}{\end{eqnarray}}
  
 \def\simge{\mathrel{%
   \rlap{\raise 0.511ex \hbox{$>$}}{\lower 0.511ex \hbox{$\sim$}}}}
\def\simle{\mathrel{
   \rlap{\raise 0.511ex \hbox{$<$}}{\lower 0.511ex \hbox{$\sim$}}}}

\begin{document}

\title{Strongly coupled $\mathcal{N}=4$ super Yang-Mills plasma on the Coulomb branch II: Transport coefficients and hard probe parameters}

\author{Kiminad A. Mamo}
\email{kiminad.mamo@stonybrook.edu}
\affiliation{Department of Physics and Astronomy, Stony Brook University, Stony Brook, New York 11794-3800, USA}
\affiliation{Department of Physics, University of Illinois, Chicago, Illinois 60607, USA}



\date{\today}
\begin{abstract}
We study $\mathcal{N} = 4$ super Yang-Mills theory on the Coulomb branch (cSYM) by using its Type IIB supergravity dual. We compute the transport coefficients, and hard probe parameters of $\mathcal{N} = 4$ cSYM at finite temperature $T$. We use the rotating black 3-brane solution of Type IIB supergravity with a single non-zero rotation parameter $r_{0}$ after analytically continuing $r_{0}\rightarrow -ir_{0}$, and in an ensemble where the Hawking temperature $T$ and a scalar condensate $<\mathcal{O}>\sim r_{0}^4$ are held fixed. We find that the bulk viscosity to entropy density ratio of the large black hole branch decreases with temperature and has a maxima around the critical temperature $T_{c}$ while, for the small black hole branch, it increases with temperature. The other transport coefficients and parameters of hard probes, such as the conductivity, jet quenching parameter, drag force, and momentum diffusion coefficients of the large black hole branch increase with temperature and asymptote to their conformal value while, for the small black hole branch, they decrease with temperature.
\end{abstract}


\maketitle

\setcounter{footnote}{0}


\section{Introduction}
The AdS/CFT correspondence \cite{Maldacena:1997re, Gubser:1998bc, Witten:1998qj} is an important tool to compute the hydrodynamic transport coefficients and hard probe parameters of a strongly coupled plasma \cite{Policastro:2002se, Iqbal:2008by, Gubser:2006bz, Herzog:2006gh, Gubser:2006nz, CasalderreySolana:2006rq, Liu:2006ug, Gubser:2008sz}.

In this paper, we use the AdS/ CFT correspondence to study a strongly coupled $\mathcal{N} = 4$ super Yang-Mills plasma on the Coulomb branch. In this branch, a scale $\Lambda$ is generated dynamically through the Higgs mechanism where the scalar particles $\Phi_{i}$ (i=1...6) of $\mathcal{N}=4$ SYM acquire a non-zero vacuum expectation value (VEV) that breaks the conformal symmetry, and the gauge symmetry $SU(N_{c})$ to its subgroup $U(1)^{N_{c}-1}$ but preserves $\mathcal{N} = 4$ supersymmetry and the gauge coupling is not renormalised \cite{Kraus:1998hv}.

The thermodynamics of $\mathcal{N} = 4$ super Yang-Mills on the Coulomb (cSYM) is investigated in some detail in \cite{Mamo:2016k}. In this paper, we will study its hydrodynamic transport coefficients and hard probes by using its dual geometry given by a rotating black 3-brane solution of Type IIB supergravity with a single non-zero rotation parameter $r_{0}$ \cite{Kraus:1998hv, Brandhuber:1999jr, Gubser:1998jb, Cvetic:1999ne,Avramis:2006ip,Caceres:2006dj}, after analytically continuing $r_{0}\rightarrow -ir_{0}$, and in an ensemble where the Hawking temperature $T$ and a scalar condensate $<\mathcal{O}>\sim r_{0}^4$ are held fixed \cite{Mamo:2016k}.

So far, the computations of the transport coefficients of the rotating black 3-brane have been limited to the grand canonical ensemble (where temperature $T$ and angular velocity $\Omega$ or chemical potential $\mu$ are held fixed), and canonical ensemble (where temperature $T$ and angular momentum density $J$ or charge density $\rho\,\,=<J^{0}>$ are held fixed), see \cite{Avramis:2006ip, Caceres:2006dj, Son:2006em, DeWolfe:2011ts}. In \cite{Cai:1998ji, Wu:2014xva}, it was shown that for planar rotating black 3-branes the two ensembles have different thermodynamics, for example, there is Hawking-Page transition in the canonical ensemble but not in the grand canonical ensemble.

The outline of this paper is as follows: In section \ref{actionIIB}, we write down the 5-dimensional Type IIB supergravity action and its rotating black 3-brane or R-charged black hole solution.

In section \ref{transport}, we compute the hydrodynamic transport coefficients, such as shear viscosity, bulk viscosity and conductivity of the rotating black 3-brane solution dual to $\mathcal{N} = 4$ super Yang-Mills on the Coulomb branch (cSYM) at strong coupling.

In section \ref{probes}, we calculate the the drag force, momentum diffusion coefficient, and jet quenching parameter on the rotating black 3-brane solution. 

\section{\label{actionIIB}Type IIB supergravity action and background solution}
The action for the $U(1)^3$ consistent truncation of Type IIB supergravity on $S^5$ is given by \cite{Cvetic:1999xp, Cvetic:2000nc}, see also \cite{Donos:2011qt, Mamo:2015aia}
\begin{equation}\label{action}
S=\frac{1}{16\pi G_{5}}\int d^5x\,\sqrt{-g_{5}}\,\mathcal{L}_{bulk}
\end{equation}
where
\begin{align}
\mathcal{L}_{bulk}&= (\mathcal{R}-V)-\frac{1}{2}\,\sum_{I=1}^{2}\,(\partial\varphi_{I})^2-\frac{1}{4}R^2\,\sum_{a=1}^{3}X_{a}^{-2}\,(F^{a})^2\,,\nonumber\\
F_{\mu\nu}^a&=\partial_{\mu}A_{\nu}^a-\partial_{\nu}A_{\mu}^a,\quad V=-\frac{4}{R^2}\,\sum_{a=1}^{3}X_{a}^{-1}\,,\notag\\
X_{1}&=e^{-\frac{1}{\sqrt{6}}\varphi_{1}-\frac{1}{\sqrt{2}}\varphi_{2}},\quad X_{2}=e^{-\frac{1}{\sqrt{6}}\varphi_{1}+\frac{1}{\sqrt{2}}\varphi_{2}},\quad X_{3}=e^{\frac{2}{\sqrt{6}}\varphi_{1}}\,.
\end{align}
We have dropped the Chern-Simons term from the action (\ref{action}) since it does not play any role in our discussion below.

In this paper, we compute the hydrodynamic and hard probe transport coefficients the following rotating black 3-brane solution of the above action (\ref{action})\cite{Behrndt:1998jd, Son:2006em}
\begin{equation}
ds_{(5)}^2 = \frac{r^2}{R^2}{H}^{1/3}\Big(-f\, dt^2+dx^2 + dy^2 + dz^2\Big)
+\frac{ H^{-2/3}}{\frac{r^2}{R^2}f}dr^2\,,
\label{metric}
\end{equation}
where
\begin{equation}
f= 1-\frac{r^4_h}{r^4}\frac{H(r_{h})}{H(r)}\,,\,\,\,H=1-\frac{r_{0}^2}{r^2}\,,
\end{equation}
\bea
\varphi_{1} &=& \frac{1}{\sqrt{6}}\ln H\,,\,\, \varphi_{2} = \frac{1}{\sqrt{2}}\ln H  \,,\nonumber\\
A^1_t &=&i\frac{r_{0}}{R^2}\frac{r_{h}^2\sqrt{H(r_{h})}}{r^2H(r)}\,,\nonumber\\
r_{h}^2 &=&\frac{1}{2}\Big(r_{0}^2+\sqrt{r_{0}^4+4m}\Big)\,,
\eea
$\kappa=\frac{r_{0}^2}{r_{h}^2}$, and $A^2_t=A^3_t=0$. Our metric (\ref{metric}) is equivalent to the metric used in \cite{Son:2006em} after analytically continuing $r_{0}\rightarrow -i\sqrt{q}$. Note that having an imaginary gauge potential doesn't lead to any inconsistencies in the 5-dimensional bulk spacetime, since all physical quantities in the bulk are given in terms of $(\partial_{r}A^1_{t})^2$. In the field theory side, having a finite imaginary chemical potential $\mu$, means that we are exploring the phase diagram of $\mathcal{N}=4$ cSYM at finite temperature $T$ and imaginary chemical potential $\mu$, see \cite{Roberge,Alford:1998sd,DElia:2002tig,deForcrand:2002hgr} for the study of the QCD phase diagram on the lattice at finite imaginary chemical potential without any inconsistencies.

\begin{figure}
 \begin{center}
\includegraphics[width=0.48\textwidth]{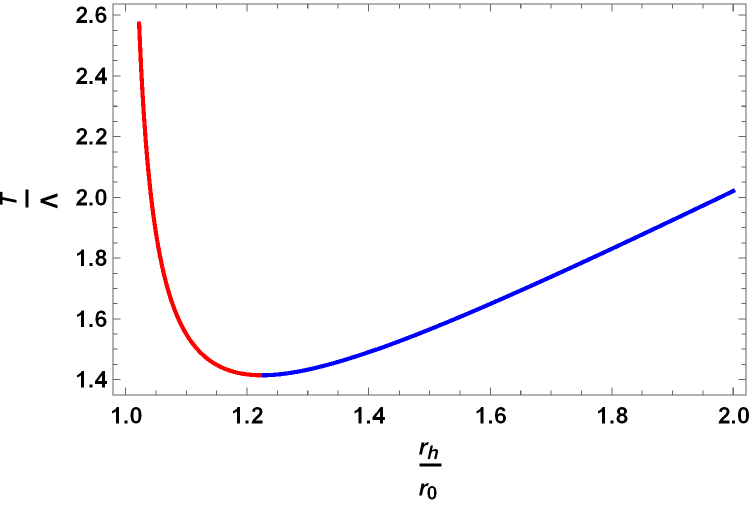}
\caption{Hawking temperature $\frac{T}{\Lambda}$ vs. the radius of the horizon $\frac{r_{h}}{r_{0}}$ (\ref{T}), normalized by the energy scale $\Lambda$ and rotation parameter $r_{0}$, respectively.}\label{fT}
 \end{center}
\end{figure}

The Hawking temperature $T$ of the rotating black 3-brane solution (\ref{metric}) is given by
\begin{equation}\label{T}
\frac{T}{\Lambda} = {1-\frac{1}{2}\kappa\over
\sqrt{\kappa-\kappa^2}}\,\,,
\end{equation}
where $T_{0}=\frac{r_{h}}{\pi R^2}$, $\Lambda=\frac{r_{0}}{\pi R^2}$, and $\kappa=\frac{r_{0}^2}{r_{h}^2}=\frac{\Lambda^2}{T_{0}^2}$. We have plotted $\frac{T}{\Lambda}$ in Fig.~\ref{fT}. We can also invert (\ref{T}) to find
\begin{equation}\label{k}
 \kappa=\frac{1+\frac{T^2}{\Lambda^2}\Big(1\mp\sqrt{\frac{T^2}{\Lambda^2}-2}\Big)}{\frac{1}{2}+2\frac{T^2}{\Lambda^2}}\,,
\end{equation}
and
\begin{equation}\label{t0}
 \frac{T_{0}^2}{T^2}=\frac{2+\frac{1}{2}\frac{\Lambda^2}{T^2}}{1+\frac{T^2}{\Lambda^2}\Big(1\mp\sqrt{\frac{T^2}{\Lambda^2}-2}\Big)}\,.
\end{equation}
Note that in (\ref{k}) and (\ref{t0}) $"-"$ corresponds to large black hole branch and $"+"$ corresponds to small black hole branch.

The entropy density $s(T,\Lambda)$, for our ensemble where $T$ and $\Lambda$ are held fixed, is given by
\begin{eqnarray}
s(T,\Lambda)&=& {A_H\over 4 G_5 V_3}=\frac{1}{4G_{5}}\sqrt{g_{xx}(r_{h})g_{yy}(r_{h})g_{zz}(r_{h})}\,\nonumber\\
&=& {\pi^2 N_c^2 T_{0}^3 \over 2}(1-\kappa)^{1/2}\,,
\label{entropy_density}
\end{eqnarray}
where $G_5=\pi R^3/2 N_c^2$, and $V_{3}$ is the three-dimensional volume. 

\section{\label{transport}Hydrodynamic transport coefficients of $\mathcal{N}=4$ cSYM plasma}
The transverse metric fluctuation $h_{xy}(t,z,r)$ decouples from other fluctuations, hence the shear viscosity for a general background metric $g_{\mu\nu}$ is given by \cite{Mamo:2012sy}
\begin{eqnarray}\label{eta1}
\eta &=&\frac{1}{16\pi G_{5}}\sqrt{g_{xx}(r_{h})g_{yy}(r_{h})g_{zz}(r_{h})}\frac{g_{xx}(r_{h})}{g_{yy}(r_{h})}\nonumber\\
&=& \frac{s}{4\pi}\frac{g_{xx}(r_{h})}{g_{yy}(r_{h})}.
\end{eqnarray}
Since, for our background metric (\ref{metric}) $g_{xx}=g_{yy}$, the shear viscosity $\eta$ of $\mathcal{N}=4$ cSYM is simply
\begin{equation}\label{eta2}
\frac{\eta}{s}=\frac{1}{4\pi}.
\end{equation}

Bulk viscosity $\zeta$ can be computed by closely following \cite{Gubser:2008sz}. To this end, we first replace $\varphi_{1}\rightarrow\frac{1}{2}\tilde{\varphi}_{1}$ followed by $\varphi_{2}\rightarrow\frac{\sqrt{3}}{2}\tilde{\varphi}_{1}$, to bring the Einstein-Maxwell-scalar part of our action (\ref{action}) in the same form as the action used in \cite{Gubser:2008sz}, i.e.,
\begin{equation}\label{lag}
(16\pi G_{5})\frac{\mathcal{L}}{\sqrt{-g_{5}}}= (\mathcal{R}-\tilde{V}(\tilde{\varphi}_{1}))-\frac{1}{2}\,(\partial\tilde{\varphi}_{1})^2+...\,,
\end{equation}
where
\begin{eqnarray}
\tilde{V}(\tilde{\varphi}_{1})&=&-\frac{4}{R^2}\Big(e^{\frac{2}{\sqrt{6}}\tilde{\varphi}_{1}}\Big(1+\frac{\kappa(1-\kappa)}{2\kappa^3}(e^{-\frac{3}{\sqrt{6}}\tilde{\varphi}_{1}}-1)^3\Big)\nonumber\\
&+&2e^{-\frac{1}{\sqrt{6}}\tilde{\varphi}_{1}}\Big)\,.
\end{eqnarray}

\begin{figure}
 \begin{center}
\includegraphics[width=0.48\textwidth]{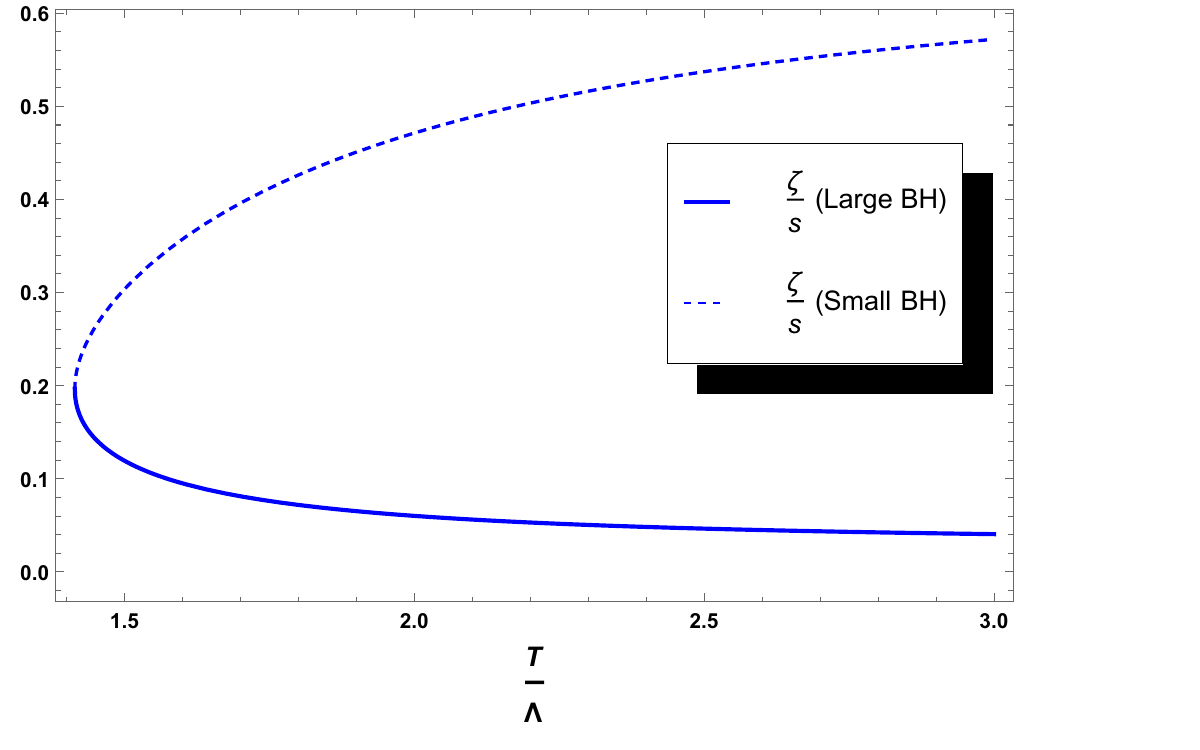}
\caption{The bulk viscosity to entropy density ratio $\frac{\zeta}{s}$ of $\mathcal{N}=4$ SYM plasma on the Coulomb branch for both large and small black holes (\ref{zeta}).}\label{fzeta}
 \end{center}
\end{figure}

In the $\tilde{r}={\varphi}_{1}(r)$ gauge, the bulk viscosity $\zeta$ up to a constant is \cite{Gubser:2008sz}
\begin{equation}\label{zeta}
  \frac{\zeta}{s}\varpropto\frac{1}{4\pi}\frac{\tilde{V}'(\tilde{r}_{h})^2}{\tilde{V}(\tilde{r}_{h})^2}\,,
\end{equation}
where $'$ denotes the derivative with respect to $\tilde{r}=\tilde{\varphi}_{1}(r)$. Note that, in the gauge $\tilde{r}=\tilde{\varphi}_{1}(r)=\frac{2}{\sqrt{6}}\ln H(r)$, the horizon of the black hole is located at $\tilde{r}=\tilde{r}_{h}=\frac{2}{\sqrt{6}}\ln H(r_h)=\frac{2}{\sqrt{6}}\ln(1-\kappa)$ where $\kappa$ is still given by (\ref{k}). We have plotted $\frac{\zeta}{s}$ in Fig.~\ref{fzeta}

\begin{figure}
 \begin{center}
\includegraphics[width=0.48\textwidth]{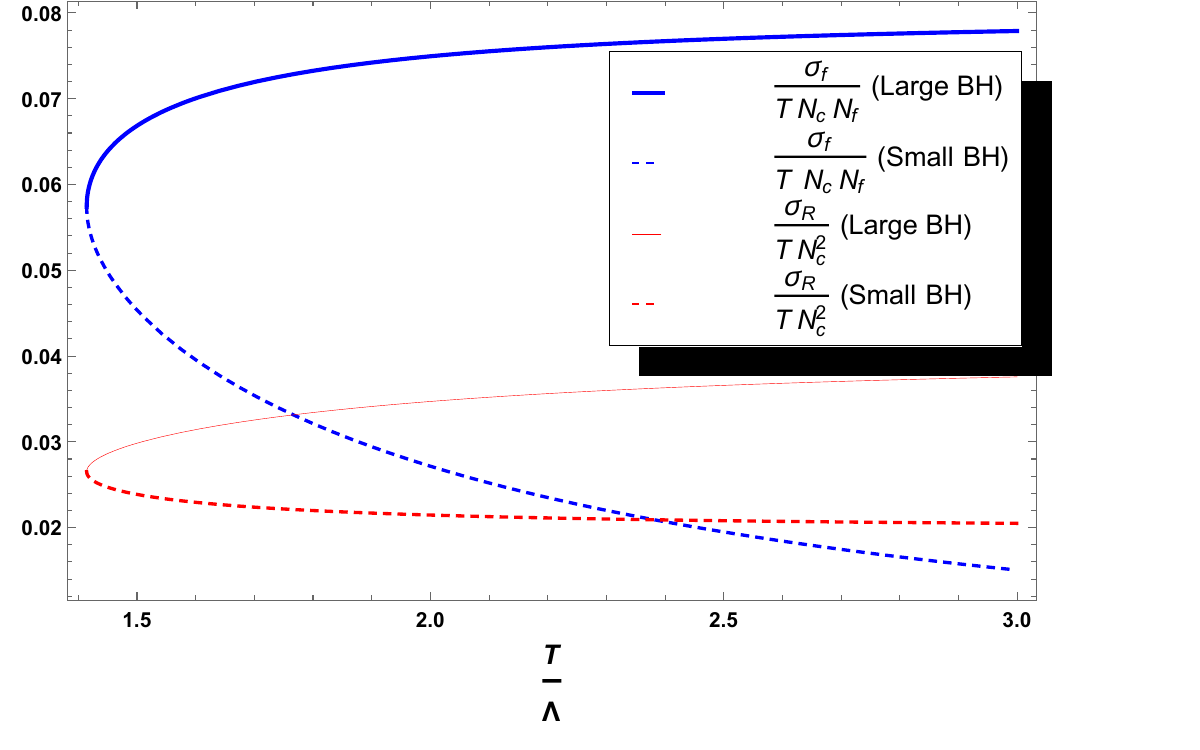}
\caption{The conductivity $\frac{\sigma_{f}}{T N_{c}N_{f}}$ of a $U(1)$ flavor charge (\ref{fsigma}), and $\frac{\sigma_{\text{R}}}{T N_{c}^2}$ of a single R-charge (\ref{rsigma}) of flavored and unflavored $\mathcal{N}=4$ SYM plasma, respectively, on the Coulomb branch for both large and small black holes.}\label{figsigma}
 \end{center}
\end{figure}

The conductivity $\sigma_{f}$ of a $U(1)$ flavor charge can simply be computed using the general formula \cite{Iqbal:2008by, Mamo:2013efa}
\begin{eqnarray}\label{fsigma}
\sigma_{f}&=&\frac{1}{ g_{5}^2}\sqrt{g_{xx}(r_{h})g_{yy}(r_{h})g_{zz}(r_{h})}g^{xx}(r_{h})\nonumber\\
&=&\frac{N_{c}N_{f}T_{0}}{4\pi}(1-\kappa)^{1/6}\,,
\end{eqnarray}
where we used $g_{5}^2=\frac{4\pi^2R}{N_{c}N_{f}}$ and a bulk $U(1)$ flavor action of the form $S_{f}=-\frac{1}{4g_{5}^2}\int d^5x\sqrt{-g}F^2$ which can be derived from the low-energy limit of the Dirac-Born-Infeld action of probe $N_{f}$ D7-branes \cite{Mateos:2007yp}. Note that there is no mixing between the gravitational and flavor gauge field fluctuations. We have plotted $\sigma_{f}$ in Fig.~\ref{figsigma}.

The conductivity $\sigma_{\text{R}}$ of a single R-charge can be computed by directly computing the two-point retarded correlation functions $G^{\mu\nu}$ of the spatial component of the R-current $J^{\mu}$, in the presence background $A^1_{t}$ which results in mixing between the gravitational and gauge field fluctuations, and using Kubo formula, i.e.,
\begin{equation}\label{rsigma}
  \sigma_{R}=\lim_{\omega\rightarrow 0}-\frac{1}{\omega}\text{Im}\,G^{xx}(\omega,\mathbf{k}=0)=\frac{N_{c}^2T_{0}}{32\pi}\frac{(2-\kappa)^2}{\sqrt{1-\kappa}}\,,
\end{equation}
where in the last line we used $G^{xx}=\frac{-i(2-\kappa)^2N_{c}^2T_{0}\omega}{32\pi\sqrt{1-\kappa}}$ which is nothing but Eq.4.34 of \cite{Son:2006em} after replacing $\kappa\rightarrow -\kappa$, and $G^{xx}\rightarrow \frac{1}{2}G^{xx}$ to compensate for the different normalisation we have for the gauge fields. We have plotted $\sigma_{R}$ in Fig.~\ref{figsigma}.

\section{\label{probes}Drag force, momentum diffusion and jet quenching in $\mathcal{N}=4$ cSYM plasma}
The Nambu-Goto (NG) action is
\be \label{NGpar}
S_{NG}=\int d\tau d\sigma \mathcal{L}(h_{ab})=-\frac{1}{2\pi\alpha'}\int d\tau d\sigma \sqrt{-det\,h_{ab}}\,,\nonumber\\
\ee
where the background induced metric on the string $h_{ab}$ is given by
\be \label{ind}
h_{ab}=g_{\mu\nu}\partial_{a}x^{\mu}(\tau,\sigma)\partial_{b}x^{\nu}(\tau,\sigma)\,.
\ee

Using the embedding $(\tau, \sigma)\Rightarrow (t(\tau, \sigma),0,0,x(\tau, \sigma), r=\sigma)$, the background induced metric $h_{ab}(\dot{z},z')$ (\ref{ind}) becomes (${\cdot} \equiv d/d\tau, {'} \equiv d/d\sigma$)
\be \label{bgindz1}
h_{ab}(\dot{x},x')=g_{tt}\partial_{a}t\partial_{b}t+g_{xx}\partial_{a}x\partial_{b}x+g_{rr}\partial_{a}r\partial_{b}r \,.
\ee
Using a particular Ansatz of the form $t(\tau,\sigma)=\tau+K(\sigma)$ and $z=v\tau+F(\sigma)$, which represents a ``trailing string'' configuration moving with velocity $v$, the background induced metric (\ref{bgindz1}) becomes \cite{Li:2016bbh}
\begin{eqnarray} \label{bgindz2}
h_{\tau\tau}(v,x')&=&g_{tt}+v^2g_{xx}\,,\nonumber\\
h_{\sigma\sigma}(v,x')&=&g_{tt}(K')^2+g_{xx}(x')^2+g_{rr}\,,\nonumber\\
h_{\tau\sigma}(v,x')&=&g_{tt} K'+g_{xx}x'v\,.
\end{eqnarray}
Finding the equation of motion from the action, we have
\be
\partial_\sigma\left({g_{tt} g_{xx}(x'-vK')\over \sqrt{-det\,h_{ab}}}\right)=0\,.
\ee

Requiring $h_{\tau\sigma}(v,x')=0$ to fix this gauge freedom, we have an additional constraint $K'=-\frac{g_{xx}}{g_{tt}}x'v$, which can be used to diagonalize (\ref{bgindz2}) as \cite{Li:2016bbh}
\begin{eqnarray} \label{bgindz3}
h_{\tau\tau}(v,x')&=&g_{tt}\Big(1+v^2\frac{g_{xx}}{g_{tt}}\Big)\,,\nonumber\\
h_{\sigma\sigma}(v,x')&=&\Big(1+v^2\frac{g_{xx}}{g_{tt}}\Big)g_{xx}(x')^2+g_{rr}\,.
\end{eqnarray}
Solving the equation of motion, in this gauge, for $x'$, we find
\be\label{sol2}
(x')^2=\frac{-C^2g_{rr}}{g^2_{xx}g_{tt}}\frac{1}{\left(1+v^2\frac{g_{xx}}{g_{tt}}\right)\big(1+\frac{C^2}{g_{tt}g_{xx}}\big)}\,.
\ee
where the integration constant $C$ is related to the conjugate momenta $\Pi=\frac{\partial\mathcal{L}}{\partial x'}=-\frac{C}{2\pi\alpha'}$. Since the factor $1+v^2\frac{g_{xx}}{g_{tt}}$ in (\ref{sol2}), for $v\neq 0$, vanishes when $-\frac{g_{tt}(r_{s})}{g_{xx}(r_{s})}=v^2$, requiring $(x')^2$ to be positive across $r=r_s$, the other factor $1+\frac{C^2}{g_{tt}g_{tt}}$ has to vanish at $r=r_{s}$ as well, which will fix the integration constant $C^2=-g_{tt}(r_{s})g_{xx}(r_s)$ for $v\neq 0$.

So, the induced metric (\ref{bgindz3}) for $v\neq 0$ becomes
\begin{eqnarray} \label{bgindz6}
h_{\tau\tau}(v,x')&=&g_{tt}\Big(1-\frac{g_{tt}(r_{s})}{g_{tt}}\frac{g_{xx}}{g_{xx}(r_{s})}\Big)\,,\nonumber\\
h_{\sigma\sigma}(v,x')&=&g_{rr}\Bigg(\frac{1}{1-\frac{g_{xx}(r_{s})g_{tt}(r_{s})}{g_{xx}g_{tt}}}\Bigg)\,.
\end{eqnarray}
which can be interpreted as a metric of a 2-dimensional black hole with a line element $ds_{(2)}^2$ given by
\be \label{2dBH}
ds_{(2)}^{2}=h_{\tau\tau}d\tau^2+h_{\sigma\sigma}d\sigma^2=-g_{tt}(-\tilde{f}(r))d\tau^2+\frac{1}{\tilde{p}(r)}d\sigma^2\,,\nonumber\\
\ee
where $\tilde{f}(r)=1-\frac{g_{tt}(r_{s})}{g_{tt}}\frac{g_{xx}}{g_{xx}(r_{s})}$, $\tilde{p}(r)=g^{rr}\big(1-\frac{g_{xx}(r_{s})g_{tt}(r_{s})}{g_{xx}g_{tt}}\big)$. The radius of the horizon $r_{s}$ of the 2-dimensional black hole is found by solving the algebraic equation $-\frac{g_{tt}(r_{s})}{g_{xx}(r_{s})}=v^2$. And, the Hawking temperature of the 2-dimensional black hole denoted as $T_{s}$ is
\be \label{Tspar}
T_{s}=\frac{1}{4\pi}\sqrt{-g_{tt}(r_{s})\tilde{f}'(r_{s})\tilde{p}'(r_{s})}\,.
\ee

\begin{figure}
 \begin{center}
\includegraphics[width=0.48\textwidth]{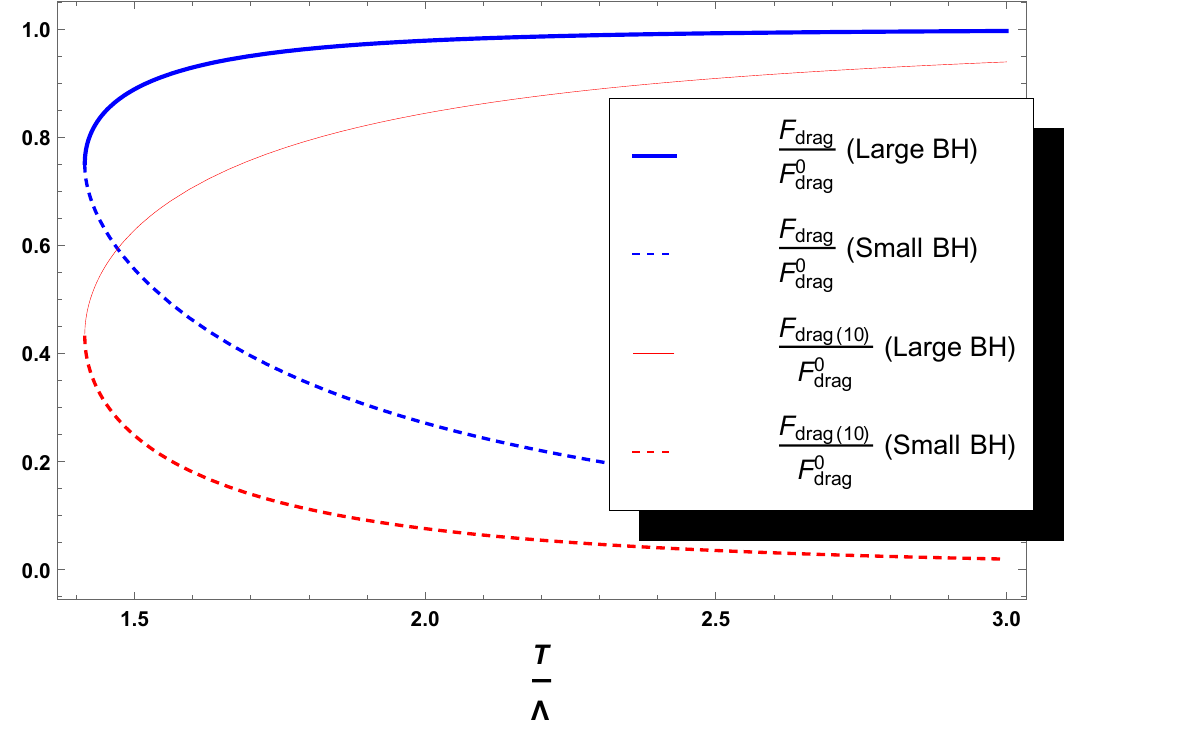}
\caption{The drag forces $\frac{F_{drag}}{F^{0}_{drag}}$ (\ref{drag}), and $\frac{F_{drag}(10)}{F^{0}_{drag}}$ of $\mathcal{N}=4$ SYM plasma on the Coulomb branch for both large and small black holes, normalized by the drag force $F^{0}_{drag}= -\frac{1}{2}\sqrt{\lambda}\pi T^2\gamma v$ of the conformal $\mathcal{N}=4$ SYM plasma \cite{Gubser:2006bz,Herzog:2006gh}.}\label{fdrag}
 \end{center}
\end{figure}

The drag force is given by \cite{Gubser:2006bz,Herzog:2006gh}, see also \cite{Li:2016bbh},
\begin{equation}\label{drag}
  F_{drag}=-\frac{C}{2\pi\alpha'}=-\frac{1}{2}\pi\sqrt{\lambda}T_{0}^2\gamma vQ(\kappa,\gamma)\,,
\end{equation}
where $Q(\kappa,\gamma)=\frac{\kappa}{2\gamma}\Big(1+\sqrt{1+4\gamma^2\frac{1-\kappa}{\kappa^2}}\Big)$, and we have used $r_{s}^2=\gamma r_{h}^2Q(\kappa,\gamma)$ which solves the algebraic equation $-\frac{g_{tt}(r_{s})}{g_{xx}(r_{s})}=v^2$. We have plotted $F_{drag}$ in Fig.~\ref{fdrag}

\begin{figure}
 \begin{center}
\includegraphics[width=0.48\textwidth]{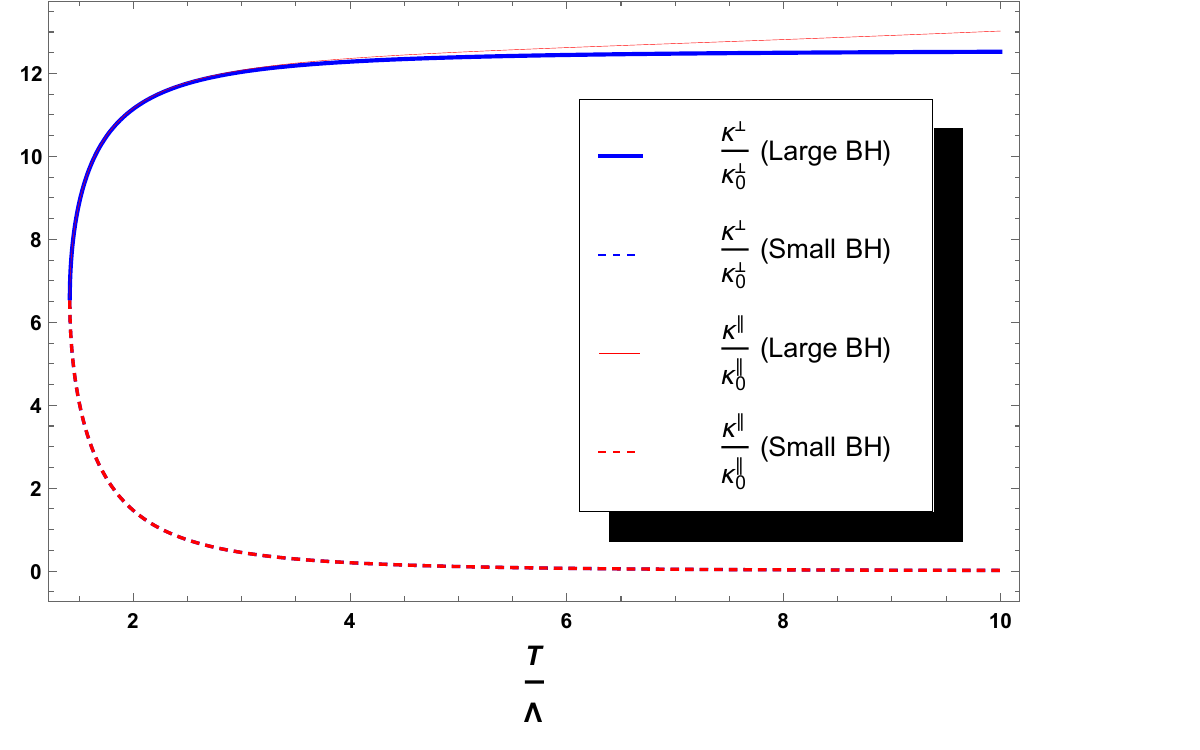}
\caption{The transverse and longitudinal momentum diffusion constants $\frac{\kappa^{\perp}(0)}{\kappa_{0}(v)}$ (\ref{dcperp}) and $\frac{\kappa^{\parallel}(0)}{\kappa_{0}(0)}$ (\ref{dcpar}), respectively, of $\mathcal{N}=4$ SYM plasma on the Coulomb branch for both large and small black holes, normalized by the momentum diffusion constant $\kappa_{0}=\kappa^{\perp}_{0}(0)=\kappa^{\parallel}_{0}(0)=\sqrt{\lambda}\pi T^3$ of the conformal $\mathcal{N}=4$ SYM plasma \cite{Gubser:2006nz,CasalderreySolana:2006rq}.}\label{fkappa}
 \end{center}
\end{figure}

The velocity dependent transverse momentum diffusion constant per unit time $\kappa^{\perp}(v)$ is given by \cite{Li:2016bbh}
\be\label{dcperp}
\kappa^{\perp}(v)=\frac{T_{s}}{\pi\alpha'}g_{xx}(r_{s})\,,
\ee
and the longitudinal momentum diffusion constant per unit time $\kappa^{\parallel}(v)$ is \cite{Finazzo:2016mhm}
\be\label{dcpar}
\kappa^{\parallel}(v)=\frac{T_{s}}{\pi\alpha'}\frac{1}{g_{xx}}\frac{(g_{tt}g_{xx})'}{(g_{tt}/g_{xx})'}\mid_{r=r_{s}}\,.
\ee
We have plotted $\kappa^{\perp}(0)$ and $\kappa^{\parallel}(0)$ in Fig.~\ref{fkappa}. Note from Fig.~\ref{fkappa} that $\kappa^{\perp}(v)\neq\kappa^{\parallel}(v)$ even at $v=0$ in $\mathcal{N}=4$ cSYM plasma, even though they are equal to each other at $v=0$ in $\mathcal{N}=4$ SYM plasma. Also note that, as can be seen in Fig.~\ref{fkappa}, the difference between $\kappa^{\perp}(v)$ and $\kappa^{\parallel}(v)$ gets enhanced with increasing $T$ and $v$.

The 5-dimensional metric (\ref{metric}) can be uplifted to the full 10-dimensional metric as \cite{Kraus:1998hv, Brandhuber:1999jr, Gubser:1998jb, Cvetic:1999ne,Avramis:2006ip,Caceres:2006dj}
\begin{eqnarray}
ds_{(10)}^2 &=& \frac{r^2}{R^2}{\tilde{H}}^{1/2}\Big(-\tilde{f}\, dt^2+dx^2 + dy^2 + dz^2\Big)
\nonumber\\
&+&R^2\Big(\tilde{H}^{1/2}d\theta^2+H\tilde{H}^{-1/2}\sin^2\theta d\phi^2\,\nonumber\\
&+&\tilde{H}^{-1/2}\cos^2\theta d\Omega_{3}^2\Big) +2A_{t}^{1}H\tilde{H}^{-1/2}R^2\sin^2\theta dtd\phi\nonumber\\
&+&\frac{\tilde{H}^{1/2}H^{-1}}{\frac{r^2}{R^2}f}dr^2\,\,,\nonumber\\
\label{10metric}
\end{eqnarray}
where
\begin{equation}\label{}
\tilde{H}=\sin^2\theta+H\cos^2\theta\,,\,\,\text{and}\,\,\,\tilde{f}=1-\frac{r_{h}^4}{r^{4}}\frac{H(r_{h})}{\tilde{H}(r)}\,,
\end{equation}
$f$ and $H$ are the same as in (\ref{metric}). Our 10-dimensional metric (\ref{10metric}) is equivalent to Eq.2.21 of \cite{Avramis:2006ip} after analytically continuing the rotation parameter $r_{0}\rightarrow -ir_{0}$, and re-writing (\ref{10metric}) in terms of $\mu\equiv m^{1/4}$. Note that the $g_{t\phi}$ component of (\ref{10metric}) is imaginary and one could make it real by analytically continuing $t\rightarrow -it$ as in \cite{Cvetic:1999rb, Brandhuber:1999jr}. However, since we are interested in real-time dynamics, such as computation of transport coefficients, we refrain from analytically continuing $t\rightarrow -it$, and we treat our 10-dimensional metric (\ref{10metric}) as a complex saddle point. Also note that $g_{t\phi}=A_{t}^1=0$ in the extremal limit $r_{h}=r_{0}$.

\begin{figure}
 \begin{center}
\includegraphics[width=0.48\textwidth]{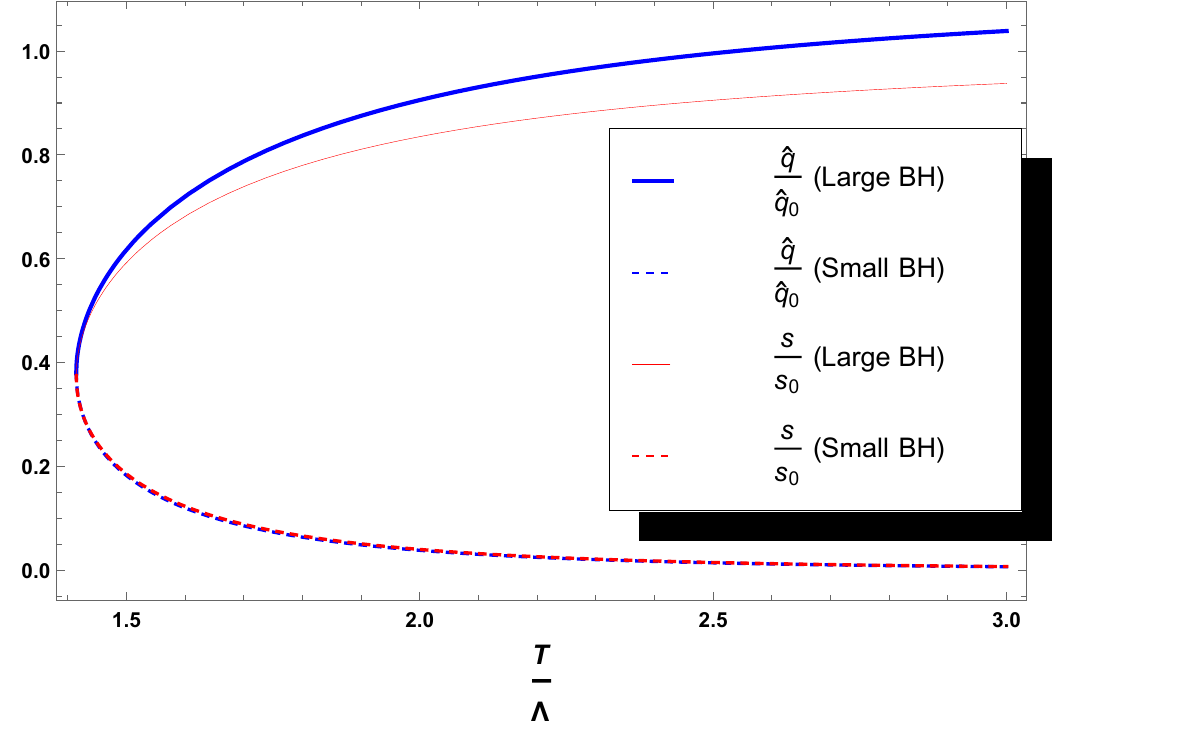}
\caption{The jet quenching parameter $\frac{\hat{q}}{\hat{q}_{0}}$ (\ref{qhat}) and entropy density $\frac{s}{s_{0}}$ of $\mathcal{N}=4$ SYM plasma on the Coulomb branch for both large and small black holes, normalized by the jet quenching parameter $\hat{q}_{0}=\frac{\pi^{3/4}\Gamma(3/4)}{\sqrt{2}\Gamma(5/4)}\sqrt{\lambda}T^3$ and entropy density $s_{0}=\frac{1}{2}\pi^2N_{c}^2T^3$ of the conformal $\mathcal{N}=4$ SYM plasma.}\label{fqhat}
 \end{center}
\end{figure}

In \cite{Caceres:2006dj} the drag force was studied using the 10-dimensiomnal metric (\ref{10metric}), and it was shown that the drag force $F_{drag(10)}$ is (shown below after re-writing it in terms of $\kappa$, and making the analytic continuation $r_{0}\rightarrow -ir_{0}$ which is equivalent to replacing $\kappa\rightarrow-\kappa$)
\begin{equation}\label{drag10}
  F_{drag(10)}= -\frac{1}{2}\sqrt{\lambda}\pi T_{0}^2\sqrt{1-\kappa}\gamma v\,.
\end{equation}
Note that (\ref{drag10}) is equivalent to the $\gamma\rightarrow\infty$ limit of (\ref{drag}), and it has similar $\sqrt{1-\kappa}$ dependence as the entropy density (\ref{entropy_density}) indicating that the drag force (\ref{drag10}) could be the measure of the color degrees of freedom of the plasma \cite{Liu:2006he}. We have plotted (\ref{drag10}) in Fig.~\ref{fdrag}.

And, in \cite{Avramis:2006ip}, it was shown that the jet quenching parameter $\hat{q}$, studied using the 10-dimensiomnal metric (\ref{10metric}), is (shown below after re-writing it in terms of $\kappa$, and making the analytic continuation $r_{0}\rightarrow -ir_{0}$ which is equivalent to replacing $\kappa\rightarrow-\kappa$)
\begin{equation}\label{qhat}
  \frac{\hat{q}}{\hat{q}_{0}}=\frac{\mathbf{K}(1/\sqrt{2})}{\mathbf{K}(n)}(2n^2)^2(2n'^2)^{1/2}\,,
\end{equation}
where $\mathbf{K}(n)$ is the complete elliptic integral of the first kind, $n^2=\frac{1-\kappa}{2-\kappa}$, $n'=\sqrt{1-n^2}$, and $\hat{q}_{0}=\frac{\pi^{3/4}\Gamma(3/4)}{\sqrt{2}\Gamma(5/4)}\sqrt{\lambda}T^3$ \cite{Liu:2006ug, Liu:2006he}. In Mathematica, the complete elliptic integral of the first kind is implemented using EllipticK[$n^2$]$\,\,\,\equiv\mathbf{K}(n)$. We have plotted $\hat{q}$ in Fig.~\ref{fqhat}.


\section{Conclusion}
We have studied the transport coefficients of the non-extremal rotating black 3-brane dual to strongly coupled $\mathcal{N}=4$ cSYM plasma, such as bulk viscosity to entropy density ratio $\frac{\zeta}{s}$ (\ref{zeta}), and conductivity $\sigma$ (\ref{fsigma})(\ref{rsigma}), see Fig.~\ref{fzeta} and Fig.~\ref{figsigma}, respectively. We have found that the bulk viscosity of the large black hole has a maxima around $T_{c}$, and its conductivity $\sigma$ asymptotes to its conformal value starting from below it. 

We have also computed the hard probe parameters of $\mathcal{N}=4$ cSYM plasma. We have shown that the drag force $F_{drag}$, momentum diffusion coefficient $\kappa$, and jet quenching parameter $\hat{q}$ increase with temperature for the large black hole, see Fig.~\ref{fdrag}, Fig.~\ref{fkappa} and Fig.~\ref{fqhat}.


\section{Acknowledgements}

The author thanks Ho-Ung Yee for stimulating discussions and helpful comments on the
draft, Krishna Rajagopal for pointing out the possible connection between jet quenching parameter and number of degrees of freedom, Pablo Morales, Andrey Sadofyev, and Yi Yin for discussions.

 \vfil


\begin{thebibliography}{99} \frenchspacing

\bibitem{Maldacena:1997re}
  J.~M.~Maldacena,
  ``The Large N limit of superconformal field theories and supergravity,''
  Adv.\ Theor.\ Math.\ Phys.\  {\bf 2}, 231 (1998)
  [hep-th/9711200].

\bibitem{Gubser:1998bc}
  S.~S.~Gubser, I.~R.~Klebanov and A.~M.~Polyakov,
  ``Gauge theory correlators from noncritical string theory,''
  Phys.\ Lett.\ B {\bf 428}, 105 (1998)
  [hep-th/9802109].

\bibitem{Witten:1998qj}
  E.~Witten,
  ``Anti-de Sitter space and holography,''
  Adv.\ Theor.\ Math.\ Phys.\  {\bf 2}, 253 (1998)
  [hep-th/9802150].

\bibitem{Policastro:2002se}
  G.~Policastro, D.~T.~Son and A.~O.~Starinets,
  ``From AdS / CFT correspondence to hydrodynamics,''
  JHEP {\bf 0209}, 043 (2002)
  [hep-th/0205052].

\bibitem{Iqbal:2008by}
  N.~Iqbal and H.~Liu,
  ``Universality of the hydrodynamic limit in AdS/CFT and the membrane paradigm,''
  Phys.\ Rev.\ D {\bf 79}, 025023 (2009)
  [arXiv:0809.3808 [hep-th]].

\bibitem{Gubser:2006bz}
  S.~S.~Gubser,
  ``Drag force in AdS/CFT,''
  Phys.\ Rev.\ D {\bf 74}, 126005 (2006)
  [hep-th/0605182].

\bibitem{Herzog:2006gh}
  C.~P.~Herzog, A.~Karch, P.~Kovtun, C.~Kozcaz and L.~G.~Yaffe,
  ``Energy loss of a heavy quark moving through N=4 supersymmetric Yang-Mills plasma,''
  JHEP {\bf 0607}, 013 (2006)
  [hep-th/0605158].

\bibitem{Gubser:2006nz}
  S.~S.~Gubser,
  ``Momentum fluctuations of heavy quarks in the gauge-string duality,''
  Nucl.\ Phys.\ B {\bf 790}, 175 (2008)
  [hep-th/0612143].

\bibitem{CasalderreySolana:2006rq}
  J.~Casalderrey-Solana and D.~Teaney,
  ``Heavy quark diffusion in strongly coupled N=4 Yang-Mills,''
  Phys.\ Rev.\ D {\bf 74}, 085012 (2006)
  [hep-ph/0605199].

\bibitem{Liu:2006ug}
  H.~Liu, K.~Rajagopal and U.~A.~Wiedemann,
  ``Calculating the jet quenching parameter from AdS/CFT,''
  Phys.\ Rev.\ Lett.\  {\bf 97}, 182301 (2006)
  [hep-ph/0605178].

\bibitem{Gubser:2008yx}
  S.~S.~Gubser, A.~Nellore, S.~S.~Pufu and F.~D.~Rocha,
  ``Thermodynamics and bulk viscosity of approximate black hole duals to finite temperature quantum chromodynamics,''
  Phys.\ Rev.\ Lett.\  {\bf 101}, 131601 (2008)
  [arXiv:0804.1950 [hep-th]].

\bibitem{Gubser:2008sz}
  S.~S.~Gubser, S.~S.~Pufu and F.~D.~Rocha,
  ``Bulk viscosity of strongly coupled plasmas with holographic duals,''
  JHEP {\bf 0808}, 085 (2008)
  [arXiv:0806.0407 [hep-th]].

\bibitem{Rougemont:2015wca}
  R.~Rougemont, A.~Ficnar, S.~Finazzo and J.~Noronha,
  ``Energy loss, equilibration, and thermodynamics of a baryon rich strongly coupled quark-gluon plasma,''
  JHEP {\bf 1604}, 102 (2016)
  [arXiv:1507.06556 [hep-th]].



\bibitem{Kraus:1998hv}
  P.~Kraus, F.~Larsen and S.~P.~Trivedi,
  ``The Coulomb branch of gauge theory from rotating branes,''
  JHEP {\bf 9903}, 003 (1999)
  [hep-th/9811120].


\bibitem{Mamo:2016k}
  K.~A.~Mamo,
  ``Strongly coupled N=4 super Yang-Mills plasma on the Coulomb branch I: Thermodynamics,''
  [arXiv:1610.09792 [hep-th]].

\bibitem{Brandhuber:1999jr}
  A.~Brandhuber and K.~Sfetsos,
  ``Wilson loops from multicenter and rotating branes, mass gaps and phase structure in gauge theories,''
  Adv.\ Theor.\ Math.\ Phys.\  {\bf 3}, 851 (1999)
  [hep-th/9906201].


\bibitem{Gubser:1998jb}
  S.~S.~Gubser,
  ``Thermodynamics of spinning D3-branes,''
  Nucl.\ Phys.\ B {\bf 551}, 667 (1999)
  [hep-th/9810225].

\bibitem{Cvetic:1999ne}
  M.~Cvetic and S.~S.~Gubser,
  ``Phases of R charged black holes, spinning branes and strongly coupled gauge theories,''
  JHEP {\bf 9904}, 024 (1999)
  [hep-th/9902195].

\bibitem{Cvetic:1999rb}
  M.~Cvetic and S.~S.~Gubser,
  ``Thermodynamic stability and phases of general spinning branes,''
  JHEP {\bf 9907}, 010 (1999)
  [hep-th/9903132].

\bibitem{Cai:1998ji}
  R.~G.~Cai and K.~S.~Soh,
  ``Critical behavior in the rotating D-branes,''
  Mod.\ Phys.\ Lett.\ A {\bf 14}, 1895 (1999)
  [hep-th/9812121].

\bibitem{Wu:2014xva}
  X.~Wu,
  ``Holographic entanglement entropy and thermodynamic instability of planar R-charged black holes,''
  Phys.\ Rev.\ D {\bf 90}, no. 6, 066008 (2014)
  [arXiv:1401.2701 [hep-th]].

\bibitem{Behrndt:1998jd}
  K.~Behrndt, M.~Cvetic and W.~A.~Sabra,
  ``Nonextreme black holes of five-dimensional N=2 AdS supergravity,''
  Nucl.\ Phys.\ B {\bf 553}, 317 (1999)
  [hep-th/9810227].
  
\bibitem{Caceres:2006dj}
  E.~Caceres and A.~Guijosa,
  ``Drag force in charged N=4 SYM plasma,''
  JHEP {\bf 0611}, 077 (2006)
  [hep-th/0605235].

\bibitem{Avramis:2006ip}
  S.~D.~Avramis and K.~Sfetsos,
  ``Supergravity and the jet quenching parameter in the presence of R-charge densities,''
  JHEP {\bf 0701}, 065 (2007)
  [hep-th/0606190].



\bibitem{Son:2006em}
  D.~T.~Son and A.~O.~Starinets,
  ``Hydrodynamics of r-charged black holes,''
  JHEP {\bf 0603}, 052 (2006)
  [hep-th/0601157].

\bibitem{DeWolfe:2011ts}
  O.~DeWolfe, S.~S.~Gubser and C.~Rosen,
  ``Dynamic critical phenomena at a holographic critical point,''
  Phys.\ Rev.\ D {\bf 84}, 126014 (2011)
  [arXiv:1108.2029 [hep-th]].





\bibitem{Cvetic:1999xp}
  M.~Cvetic {\it et al.},
  ``Embedding AdS black holes in ten-dimensions and eleven-dimensions,''
  Nucl.\ Phys.\ B {\bf 558}, 96 (1999)
  [hep-th/9903214].

\bibitem{Cvetic:2000nc}
  M.~Cvetic, H.~Lu, C.~N.~Pope, A.~Sadrzadeh and T.~A.~Tran,
  ``Consistent SO(6) reduction of type IIB supergravity on S**5,''
  Nucl.\ Phys.\ B {\bf 586}, 275 (2000)
  [hep-th/0003103].

\bibitem{Donos:2011qt}
  A.~Donos, J.~P.~Gauntlett and C.~Pantelidou,
  ``Spatially modulated instabilities of magnetic black branes,''
  JHEP {\bf 1201}, 061 (2012)
  [arXiv:1109.0471 [hep-th]].

\bibitem{Mamo:2015aia}
  K.~A.~Mamo and H.~U.~Yee,
  ``Thermalization of Quark-Gluon Plasma in Magnetic Field at Strong Coupling,''
  Phys.\ Rev.\ D {\bf 92}, no. 10, 105005 (2015)
  [arXiv:1505.01183 [hep-ph]].



\bibitem{Natsuume:2014sfa}
  M.~Natsuume,
  ``AdS/CFT Duality User Guide,''
  Lect.\ Notes Phys.\  {\bf 903}, pp.1 (2015)
  [arXiv:1409.3575 [hep-th]].



\bibitem{Mamo:2012sy}
  K.~A.~Mamo,
  ``Holographic RG flow of the shear viscosity to entropy density ratio in strongly coupled anisotropic plasma,''
  JHEP {\bf 1210}, 070 (2012)
  [arXiv:1205.1797 [hep-th]].



\bibitem{Mamo:2013efa}
  K.~A.~Mamo,
  ``Enhanced thermal photon and dilepton production in strongly coupled $N$ = 4 SYM plasma in strong magnetic field,''
  JHEP {\bf 1308}, 083 (2013)
  [arXiv:1210.7428 [hep-th]].

\bibitem{Mateos:2007yp}
  D.~Mateos and L.~Patino,
  ``Bright branes for strongly coupled plasmas,''
  JHEP {\bf 0711}, 025 (2007)
  [arXiv:0709.2168 [hep-th]].

\bibitem{Li:2016bbh}
  S.~Li, K.~A.~Mamo and H.~U.~Yee,
  ``Jet quenching parameter of the quark-gluon plasma in a strong magnetic field: Perturbative QCD and AdS/CFT correspondence,''
  Phys.\ Rev.\ D {\bf 94}, no. 8, 085016 (2016)
  [arXiv:1605.00188 [hep-ph]].


\bibitem{Liu:2006he}
  H.~Liu, K.~Rajagopal and U.~A.~Wiedemann,
  ``Wilson loops in heavy ion collisions and their calculation in AdS/CFT,''
  JHEP {\bf 0703}, 066 (2007)
  [hep-ph/0612168].

\bibitem{Finazzo:2016mhm}
  S.~I.~Finazzo, R.~Critelli, R.~Rougemont and J.~Noronha,
  ``Momentum transport in strongly coupled anisotropic plasmas in the presence of strong magnetic fields,''
  Phys.\ Rev.\ D {\bf 94}, no. 5, 054020 (2016)
  [arXiv:1605.06061 [hep-ph]].




\bibitem{Roberge}
  Andre Roberge, Nathan Weiss,
  ``Gauge Theories With Imaginary Chemical Potential and the Phases of {QCD},''
  Nucl.\ Phys.\ B {\bf 275}, 734 (1986).
 
\bibitem{Alford:1998sd} 
  M.~G.~Alford, A.~Kapustin and F.~Wilczek,
  ``Imaginary chemical potential and finite fermion density on the lattice,''
  Phys.\ Rev.\ D {\bf 59}, 054502 (1999)
  [hep-lat/9807039].

\bibitem{DElia:2002tig} 
  M.~D'Elia and M.~P.~Lombardo,
  ``Finite density QCD via imaginary chemical potential,''
  Phys.\ Rev.\ D {\bf 67}, 014505 (2003)
  [hep-lat/0209146].

\bibitem{deForcrand:2002hgr} 
  P.~de Forcrand and O.~Philipsen,
  ``The QCD phase diagram for small densities from imaginary chemical potential,''
  Nucl.\ Phys.\ B {\bf 642}, 290 (2002)
  [hep-lat/0205016].


\end{thebibliography}
\end{document}